\documentclass[aps,reprint,prb,showpacs]{revtex4-2}
\usepackage{graphicx}
\usepackage{color}
\usepackage{amsmath}
\usepackage{amssymb}
\usepackage{subcaption}
\usepackage{tabularray}
\usepackage{hyperref}

\newcommand{\inb}[3]{{\left#1 #2 \right#3}}
\newcommand{\reff}[1]{{Fig.\ref{fig:#1}}}
\newcommand{\refe}[1]{{Eq.\ref{eq:#1}}}

\newcommand{\squad}[0]{\ \ }

\begin{document}

\title{Spin Quantum Hall Effect: the Critical Exponents}
\author{Hrant Topchyan$^{1}$, Win Nuding$^{2}$,
    Andreas Kl\"umper$^{2}$, and Ara Sedrakyan$^{1}$}
\affiliation{
$^1$ A. Alikhanyan National Science Laboratory, Br. Alikhanian 2, 0036 Yerevan, Armenia\\
$^2$ Wuppertal University, Gaußstraße 20, 42119 Wuppertal, Germany}

\begin{abstract}
The spin quantum Hall effect (SQHE) provides one of the few examples of an Anderson localization transition for which exact critical exponents are known, making it an important testing ground for theories of disordered topological systems and conformal field theory. The corresponding network model, obtained by replacing the random $U(1)$ phases of the Chalker--Coddington model with random $SU(2)$ matrices, belongs to symmetry class C of the Altland--Zirnbauer classification and is believed to describe quasiparticle transport in two-dimensional disordered superconductors with broken time-reversal symmetry. In this work, we perform high-precision numerical calculations of the localization-length exponent ($\nu$) and the boundary critical exponent ($\mu$) for the SQHE network model using the recently developed $S$-matrix approach to random networks.
\end{abstract}
	\date{{\today}}

\maketitle

\section{Introduction}

The integer quantum Hall effect (IQHE) is one of the most important discoveries in condensed matter physics because it fundamentally changed our understanding of quantum phases of matter, disorder, and topology. Its impact extends well beyond semiconductor physics into modern fields such as topological insulators, topological superconductors, quantum computing, and localization theory. Perhaps the deepest consequence of the IQHE is that it introduced the concept of topological order into non-interacting electronic systems.

The IQHE provided the first clear example in which disorder is essential rather than detrimental. Impurities localize most electronic states, while only a few critical energies support extended states. As the Fermi level crosses one of these extended states, the Hall conductance changes from one plateau to the next. This led to the modern theory of Anderson localization, the scaling theory of localization, and the study of quantum critical phenomena in disordered systems.

The IQHE motivated the development of the Chalker--Coddington (CC) network model
\cite{Chalker1988}, which has become the standard description of localization--delocalization transitions in two-dimensional systems. The model describes electrons drifting along equipotential contours and tunneling at saddle points of the periodic potential, where the randomness of the potential is mimicked solely by the randomness of the distances between saddle points. This randomness generates random $U(1)$ phases for the electrons through the Aharonov--Bohm effect in a constant magnetic field. In a series of papers \cite{Zirnbauer-1997}, Zirnbauer \emph{et al.} argued that the critical point of the CC network model is described by a supersymmetric $PSL(2|2)$ Wess--Zumino--Witten (WZW) conformal field theory. Within this framework, the plateau transition of the integer quantum Hall effect is governed by a nonunitary supersymmetric critical theory possessing exact conformal invariance.

Numerical calculations of the localization-length exponent $\nu$ of the CC model have been carried out in various studies \cite{Slevin-2009,Sedrakyan-2011,Dahlhaus-2011,Fulga-2011,Obuse-2012,Sedrakyan-2015}, yielding the value $\nu \simeq 2.56 \pm 0.02$, which is not consistent with the experimental value $\nu = 2.38 \pm 0.06$ \cite{Tsui-2009}. In \cite{Sedrakyan-2017,Sedrakyan-2019,Topchyan-2024}, it was argued that the origin of this discrepancy is the neglect of saddle-point network disorder, which should be taken into account. Incorporating this type of disorder changes the localization-length exponent to $\nu = 2.372 \pm 0.017$.

The Harris criterion \cite{Harris-1974} states that uncorrelated quenched disorder is irrelevant at a clean critical point if $\nu d > 2$, where $d$ is the spatial dimension. This suggests that geometric disorder of the random lattice should also be irrelevant. Later, Luck \cite{Luck-1993} (see also \cite{Barghathi-2014,Barghathi-2016}) argued that the Harris criterion must be modified in the presence of geometric disorder. In \cite{Sedrakyan-2025}, it was demonstrated that the particular type of geometric disorder introduced in the IQHE is indeed relevant and that Luck's modification of the Harris criterion is essential in this case. As a consequence, the critical exponent $\nu$ is allowed to change.

The CC model has served as the prototype for numerous generalized network models describing the spin quantum Hall effect (SQHE), the thermal quantum Hall effect, topological superconductors, and other systems belonging to the various symmetry classes of the Altland--Zirnbauer classification \cite{Altland-1997}.

The SQHE is the spin analogue of the integer quantum Hall effect (IQHE), but it appears in systems where the charge Hall conductance may vanish while the spin Hall conductance remains quantized. It is usually discussed in the context of two-dimensional disordered superconductors with spin-rotation symmetry but broken time-reversal symmetry, corresponding to symmetry class $C$ of the Altland--Zirnbauer classification \cite{Altland-1997}. Such superconductors can exhibit quantum Hall transitions in which either the spin (class $C$) \cite{Kagalovsky1999,Senthil-1999} or the thermal (class $D$) \cite{Senthil-2000} conductivity of quasiparticles undergoes rapid transitions between quantized plateaus.

\begin{figure}
    \centering
    \includegraphics[width=\linewidth]{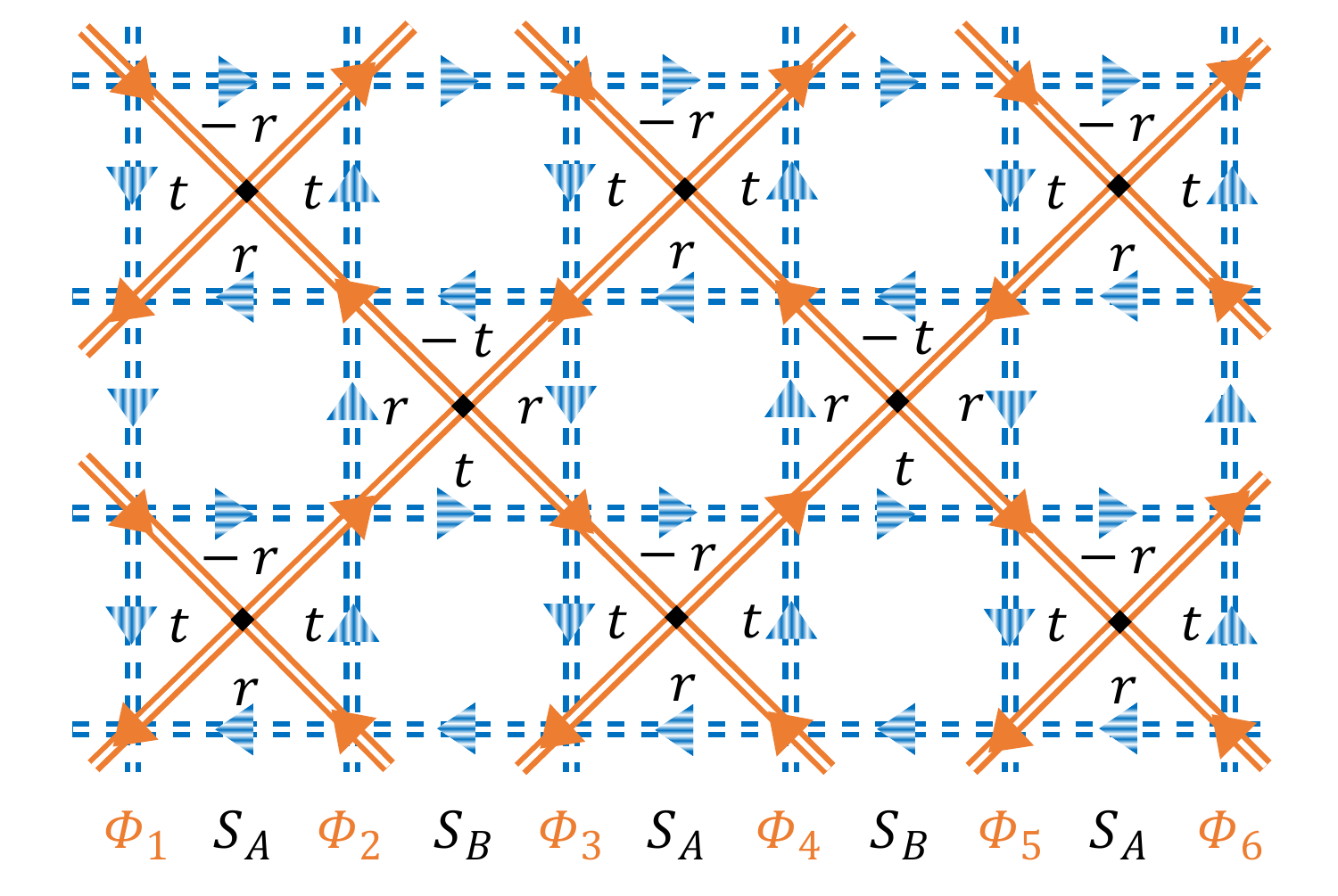}
    \caption{The scattering network underlying the spin quantum Hall plateau transition.}
    \label{fig:lattice}
    \vspace{-.5cm}
\end{figure}

The presence of spin in the SQHE naturally requires that the spin group $SU(2)$ plays the same role as the random $U(1)$ phases in the CC model \cite{Chalker1988}, which arise from the Aharonov--Bohm effect for electrons in a constant magnetic field. By analogy with the CC model, a network model for the SQHE with random $SU(2)$ group elements was formulated in \cite{Kagalovsky1999,Gruzberg1999}. It is worth noting that a model of spin fermions on the same network (see \reff{lattice}) had already been introduced earlier \cite{Sedrakyan-1986,Sedrakyan-1987} in connection with the so-called sign-factor problem in the three-dimensional gauge Ising model.

Numerical studies of the spin quantum Hall transition were reported in \cite{Kagalovsky1999,Gruzberg1999,Subramaniam2008}. In addition, the localization-length exponent was shown analytically to take the exact value $\nu = 4/3$ \cite{Gruzberg1999,Subramaniam2008}. In contrast, the boundary critical exponent $\mu$ was initially conjectured to be $\mu = 3/2$ \cite{Kagalovsky1999} and was supported by early numerical results \cite{Gruzberg1999}, but was later shown to have the exact value $\mu = 8/7$ through an analysis of boundary criticality and multifractality \cite{Subramaniam2008}.

In this paper, we determine numerically the localization-length exponent ($\nu$) and the boundary critical exponent ($\mu$) for the spin quantum Hall transition using the $S$-matrix approach to network models developed in \cite{Sedrakyan-2025}. We restrict our analysis to the regular network and achieve a numerical precision of the order of $10^{-3}$. We expect that these high-precision estimates of the critical exponents will provide useful benchmarks for identifying the conformal field theory describing the critical point of the spin quantum Hall transition.

\section{The model}

The ordinary CC model describes the motion of a single particle in a disordered potential
in the presence of a magnetic field \cite{Chalker1988}.
The particles are allowed to move along the boundaries of the Fermi-sea puddles of a Landau level
that emerge in the disordered potential, thereby acquiring an Aharonov--Bohm phase.
They are also allowed to tunnel between neighboring puddles
in the vicinity of the saddle points of the potential.
The CC model is a scattering network defined on a regular square lattice with phase disorder.
Each node of the network represents a $2 \times 2$ scattering matrix
\begin{equation}
    s=\begin{pmatrix}\ \ \mathfrak{r} & \mathfrak{t} \\
    -\mathfrak{t} & \mathfrak{r} \end{pmatrix} \squad,
\end{equation}
which maps the two incoming amplitudes $(\psi_{I,1},\psi_{I,2})$
to the outgoing amplitudes $(\psi_{O,1},\psi_{O,2})$. Here,
$\mathfrak{r}$ and $\mathfrak{t}$ denote the reflection and tunneling amplitudes, respectively.
Disorder is incorporated through random phases $\varphi$
defined on the links of the network.

The CC model can be generalized such that the propagating particles
transform in a representation of an arbitrary group.
For the SQHE, we are interested in the two-component spin-$1/2$ representation
of the group $SU(2)$.
The states $\psi$ are then two-component spinors $(\psi_\uparrow,\psi_\downarrow)$.
The scattering amplitudes $\mathfrak{r}$ and $\mathfrak{t}$ themselves become $2\times2$ matrices.
However, scattering at the nodes does not change the spin,
so that $\mathfrak{r}$ and $\mathfrak{t}$ are diagonal,
\begin{equation}
\mathfrak{r}=\text{diag}(r_\uparrow,r_\downarrow)\squad,\squad
\mathfrak{t}=\text{diag}(t_\uparrow,t_\downarrow)\squad.
\end{equation}
The amplitudes $r_\sigma$ and $t_\sigma$,
with $\sigma\in\{\uparrow,\downarrow\}$,
depend on the energies $x_\sigma$ of the corresponding channels relative to the saddle-point energy
of the potential. The parameters $x_\sigma$ are taken to be uniform throughout the system:
\begin{equation}
r_\sigma=(1+e^{2x_\sigma})^{-1/2} \squad,\squad
t_\sigma=(1+e^{-2x_\sigma})^{-1/2} \squad.
\end{equation}
For later convenience,
we introduce the average energy $\varepsilon=(x_\uparrow+x_\downarrow)/2$ and
the separation $\Delta=x_\uparrow-x_\downarrow$.
The physical properties of the system are invariant under the transformations
$\Delta\rightarrow-\Delta$ and $\varepsilon\rightarrow-\varepsilon$.
The first transformation is equivalent to interchanging the channels
$\uparrow\leftrightarrow\downarrow$.
Applying both transformations is equivalent to interchanging
$\mathfrak{r}\leftrightarrow\mathfrak{t}$,
which in turn corresponds to a translation of the lattice.

The phases $\varphi$ are likewise promoted to uniformly distributed random $SU(2)$ matrices.
This is achieved by choosing
\begin{equation}
\varphi =
\begin{pmatrix}
a+bi & c+di \\
-c+di & a-bi
\end{pmatrix},
\end{equation}
where $a,b,c,d\in\mathbb{R}$ form a random unit vector $(a,b,c,d)$
uniformly distributed on the three-sphere $S^3$.
Such vectors can be generated by drawing $a_0,b_0,c_0,d_0$
from a Gaussian distribution, which ensures isotropy in $\mathbb{R}^4$,
and subsequently normalizing the resulting vector to unit length.

\begin{figure}
\centering
\begin{subfigure}[l]{.7\linewidth}
\includegraphics[width=\textwidth]{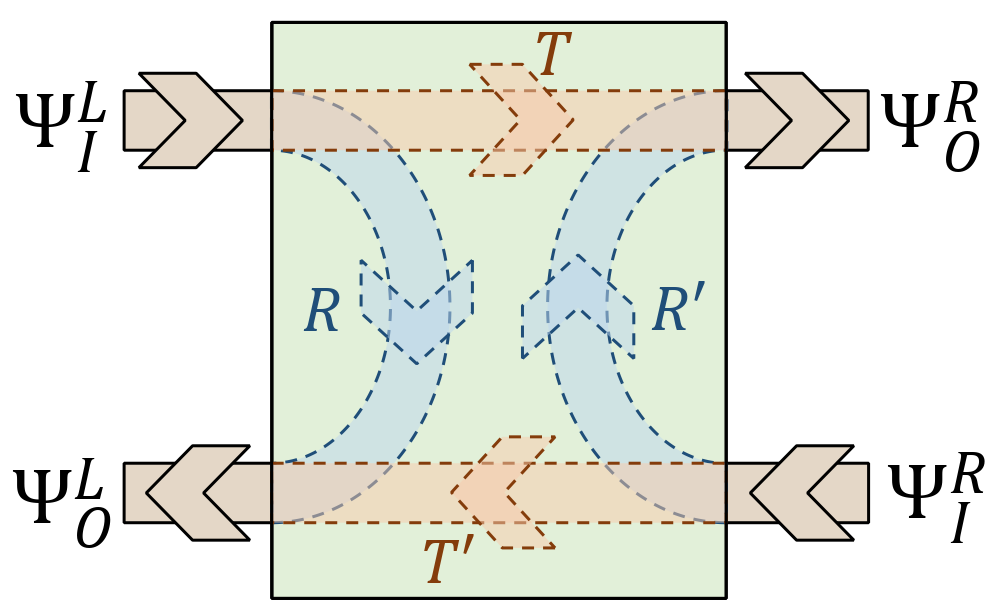}
\caption{Structure of a single $S$ matrix,
mapping $(\Psi_{I,L},\Psi_{I,R})$ to
$(\Psi_{O,L},\Psi_{O,R})$.}
\label{fig:smat_def}
\end{subfigure}
\begin{subfigure}[l]{.8\linewidth}
\includegraphics[width=\textwidth]{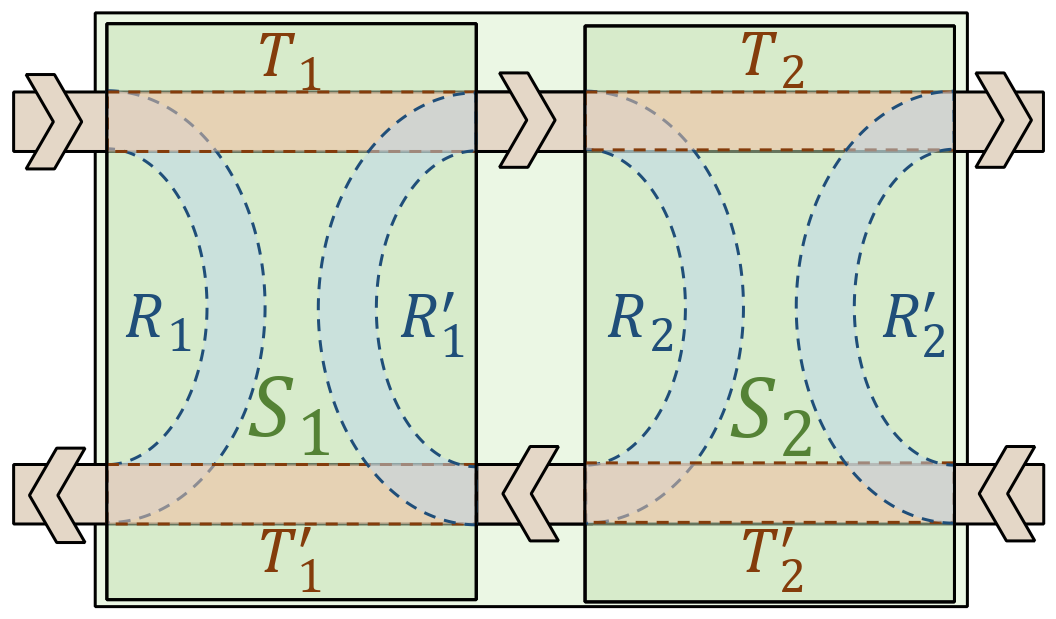}
\caption{Amalgamation of two $S$ matrices, $S_1\star S_2$,
including the scattering loop $R_2R_1'$.}
\label{fig:smat_mul}
\end{subfigure}
\caption{Schematic representation of the structure of the $S$ matrices.
The two reflection blocks, $R$ and $R'$ (blue stripes), and
the two transmission blocks, $T$ and $T'$ (red stripes), are shown,
with the arrows indicating the direction of the mapping.}
\end{figure}

We study the localization-length exponent by analyzing the finite-size scaling
of the localization length.
The localization length itself is determined from the particle transport
through a given system.
A system of width $M$ and length $L$, where the latter denotes the transport direction,
can be represented as an amalgamation of $2L$ layers of composite scattering matrices $S$,
\begin{equation}
S=\begin{pmatrix} R & T' \\ T & R'\end{pmatrix} :
\begin{pmatrix} \Psi_{I,L} \ \Psi_{I,R} \end{pmatrix} \rightarrow
\begin{pmatrix} \Psi_{O,L} \ \Psi_{O,R} \end{pmatrix}.
\end{equation}
Here, $\Psi_{I/O,L/R}$ denote the sets of incoming ($I$) and outgoing ($O$) amplitudes
on the left ($L$) and right ($R$) sides of a given layer,
while $R$, $R'$, $T$, and $T'$ denote the reflection and transmission blocks
of the matrix $S$ (see \reff{smat_def}).
Each $\Psi$ consists of $M$ spinors $\psi$, and each spinor has two components,
$\uparrow$ and $\downarrow$, so that $S$ is a $4M\times4M$ matrix.
The elements of $S$ are constructed from the elementary scattering matrix $s$.

The total scattering matrix takes the form
\begin{equation}
\mathcal{S}=S_A\star\Phi_1\star S_B\star\Phi_2\star S_A
\star\cdots\star S_B\star\Phi_{2L}\squad.
\end{equation}
Here, the matrices $\Phi_i$ are block-diagonal compositions
of the random matrices $\varphi$ defined on the links of the network.
The labels $A$ and $B$ on the matrices $S$ indicate that, owing to the structure
of the network, the composite scattering matrices have slightly different arrangements
and occur in a staggered sequence.
The operation ``$\star$'' denotes the amalgamation of two scattering matrices.
The blocks of the resulting matrix $S_{1,2}=S_1\star S_2$
are determined by the nontrivial composition rules \cite{S-matrix}
\begin{equation}
\begin{split}
R_{1,2} &= R_1 + T_1'R_2\Sigma'T_1
\squad,\squad
T'_{1,2} = T_1'\Sigma T_2'
\squad,\\
T_{1,2} &= T_2\Sigma'T_1
\squad,\squad
R'_{1,2} = R_2' + T_2R_1'\Sigma T_2'
\squad.
\end{split}
\end{equation}
Here,
\begin{equation}
\begin{split}
\Sigma  &\equiv (1-R_2R_1')^{-1}
= 1+R_2\Sigma'R_1
\squad,\\
\Sigma' &\equiv (1-R_1'R_2)^{-1}
= 1+R_1'\Sigma R_2
\end{split}
\end{equation}
encapsulate all contributions from repeated scattering around the loop
shown in \reff{smat_mul}.
For the matrices $\Phi_i$, the blocks $R$ and $R'$ vanish,
which simplifies the composition rules.

\section{Calculations}

The correlation length $\xi$ for a right-moving state can be obtained from the relation
$\inb<{ \inb|{\Psi_{I,L}^{(0)} \Psi_{O,R}^{(x)}}| }>\propto e^{-x/\xi}$,
where the superscript denotes the position along the direction of transport.
If we denote the blocks of $\mathcal S$ by $\mathcal R, \mathcal R', \mathcal T, \mathcal T'$,
then $\xi$ is given by
\begin{equation}
\gamma \equiv \xi^{-1} = -\frac{\ln [\text{tr } \mathcal T^\dagger\mathcal T]}{2L} \squad.
\end{equation}
For sufficiently large $L$, the trace of $\mathcal T^\dagger\mathcal T$
is dominated by its largest eigenvalue, since all other eigenvalues are exponentially suppressed.
Thus, $\gamma$ is equivalent to the leading Lyapunov exponent $\gamma_0$.
An analogous statement applies to a left-moving state, with $\mathcal T'$ replacing $\mathcal T$.

It is convenient to introduce the dimensionless scaled Lyapunov exponent
$\Gamma=M\gamma$.
Taking into account the symmetry under
$\varepsilon\rightarrow-\varepsilon$ and $\Delta\rightarrow-\Delta$,
the finite-size scaling behavior of $\Gamma$
near the critical point $\varepsilon=\Delta=0$ must take the form
\begin{equation}
\Gamma(\varepsilon, \Delta, M) =
F(\varepsilon \cdot M^{1/\nu}, \Delta \cdot M^{1/\mu}, M^y)
\end{equation}
where $\nu$ and $\mu$ are the scaling exponents along the two principal directions,
and $y$ is the leading irrelevant exponent.
In the fitting procedure, we use the truncated expansion
\begin{equation}
\begin{split}
\Gamma(\varepsilon, \Delta, M) \approx F_0+k\cdot M^y \hspace{7pt}& \\
+ F_{20}\cdot \varepsilon^2 \cdot M^{2/\nu} &+ F_{02}\cdot \Delta^2 \cdot M^{2/\mu} \\
+ F_{40}\cdot \varepsilon^4 \cdot M^{4/\nu} &+ F_{04}\cdot \Delta^4 \cdot M^{4/\mu} \\
&+F_{06}\cdot \Delta^6 \cdot M^{6/\mu}
\squad,
\end{split}
\label{eq:scaling}
\end{equation}
which is chosen to minimize the confidence bounds of the parameters of interest,
namely $\nu$, $\mu$ and $y$.
The values are consistent throughout various reasonable truncation choices.

\begin{figure}
\begin{subfigure}[l]{.49\linewidth}
\includegraphics[width=\textwidth]{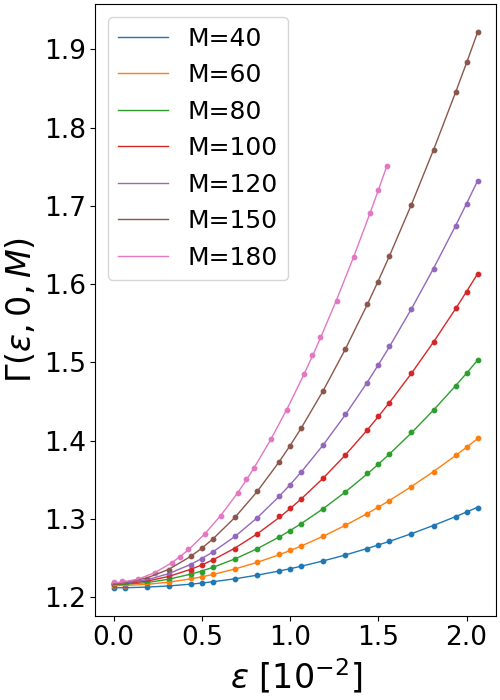}
\caption{Scaling behavior along the slice $\Delta=0$.
    $\chi^2/\text{dof}=2.5$.}
\label{fig:E}
\end{subfigure}
\begin{subfigure}[l]{.49\linewidth}
\includegraphics[width=\textwidth]{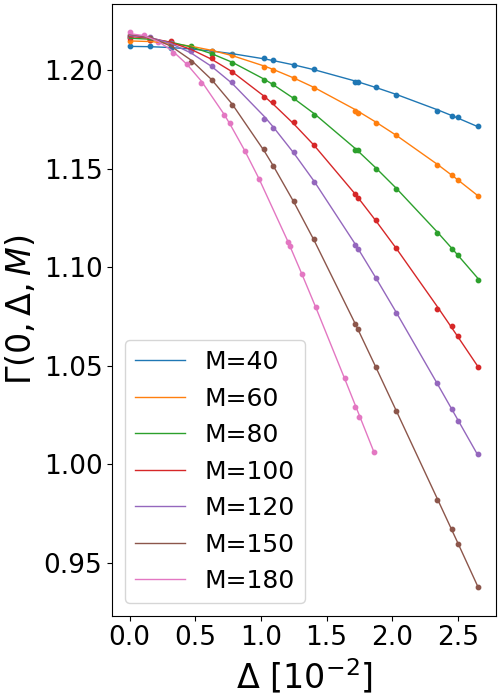}
\caption{Scaling behavior along the slice $\varepsilon=0$.
    $\chi^2/\text{dof}=2.0$.}
\label{fig:D}
\end{subfigure}
\begin{subfigure}[l]{\linewidth}
\includegraphics[width=\textwidth]{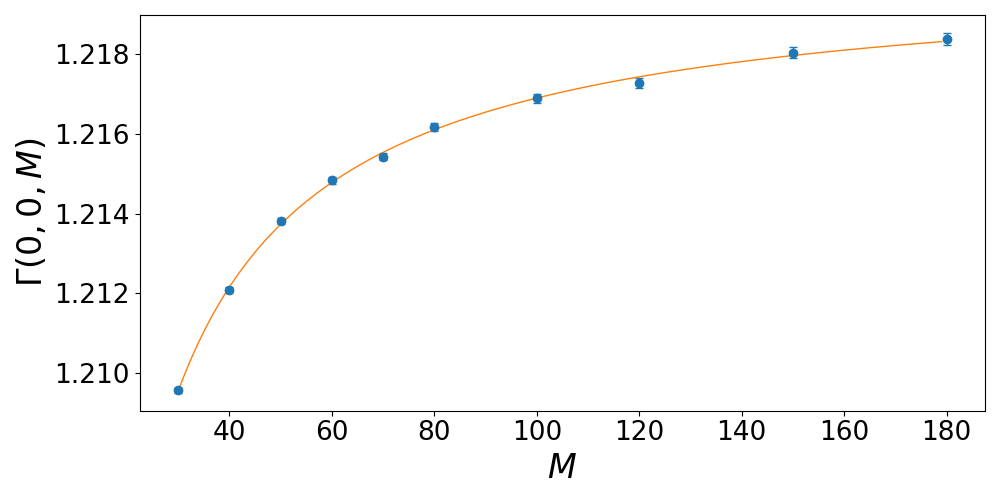}
\caption{Scaling behavior at the critical point $\varepsilon=\Delta=0$.
    $\chi^2/\text{dof}=1.0$.}
\label{fig:y}
\end{subfigure}
\caption{Computed values of $\Gamma(\varepsilon,\Delta,M)$ (dots)
and fitting curves obtained from the finite-size scaling ansatz
used to determine the critical exponents.}
\label{fig:data}
\end{figure}

To extract the exponents $\nu$, $\mu$, and $y$,
we perform three series of calculations of the scaled Lyapunov exponent $\Gamma$
for systems of size $M\times L$ and different widths $M$.
We use between $8$ and $10$ values of $M$ in the range $[30,180]$
and fix $2L=5\cdot 10^6$.
To determine $\nu$, we set $\Delta=0$ and vary $\varepsilon$
over $22$ values in the range $[0,0.02]$.
To determine $\mu$, we set $\varepsilon=0$ and vary $\Delta$
over $17$ values in the range $[0,0.025]$.
Finally, to determine $y$, we set $\varepsilon=\Delta=0$.
Each value of $\Gamma$ is obtained by averaging over $160$ disorder realizations,
except at $\varepsilon=\Delta=0$,
where an enlarged ensemble of $1600$ realizations is used.
The variances $\delta \Gamma$ are estimated from the standard deviations of $\Gamma$
across different realizations.

The resulting data are shown in \reff{data}.
First, the value
$y=-0.98(4)\approx-1$
is obtained from the truncated scaling form \refe{scaling} using the data at the critical point
(\reff{y}).
The values of $\nu$ and $\mu$ are then determined from the same expansion,
with the previously obtained value fixed to $y=-1$
(\reff{E} and \reff{D}), yielding
\begin{equation}
\nu=1.331(3) \squad,\squad \mu=1.151(2) \squad.
\end{equation}

The resulting values are inconsistent with the original estimates reported in
\cite{Kagalovsky1999},
namely $\nu=1.12$ and $\mu=1.45$.
However, our value of $\nu$ agrees with the analytical result
$\nu=4/3$ obtained in \cite{Gruzberg1999}.
In \cite{Gruzberg1999}, the value of $\mu$ was tentatively identified as
$\mu=3/2$, in agreement with \cite{Kagalovsky1999}.
It was later revised to
$\mu=8/7\approx1.143$ in \cite{Subramaniam2008}.
This exact value is only marginally different from
the value $\mu=1.151(2)$ found in the present study.

\section{Conclusion}

In this work, we have investigated the critical properties of the spin quantum Hall transition using the $S$-matrix approach to network models developed in \cite{Sedrakyan-2025}. Restricting our analysis to the regular network, we have determined the localization-length exponent $\nu$ and the boundary critical exponent $\mu$ with a numerical precision of order $10^{-3}$.

Our result $\nu=1.331(3)$ is consistent with the exact value $\nu=4/3$ obtained from the mapping of the class-$C$ network model to classical percolation. For the boundary critical exponent, we obtain $\mu=1.151(2)$, which is close to, but statistically inconsistent with, the exact prediction $\mu=8/7$. While our result for $\nu$ confirms the accuracy of the $S$-matrix method, the small discrepancy found for $\mu$ indicates that residual finite-size or systematic effects may still be relevant.

This work provides a high-precision benchmark for the critical behavior of the regular spin quantum Hall network. Such benchmark calculations are important for future investigations of geometrically disordered networks, for which the applicability of the Harris--Luck criterion and the possibility of changes in the universality class remain open questions. Extending the $S$-matrix approach to random network geometries for the SQHE is therefore a natural direction for future research.

Finally, the high-precision values of the critical exponents reported here provide valuable constraints on candidate conformal field theories describing the spin quantum Hall critical point. In particular, they may help to test and distinguish between proposed supersymmetric conformal field theories based on Wess--Zumino--Witten models for symmetry class $C$.

\section*{Acknowledgments}
The authors thank John Chalker and Ilya Gruzberg for insightful discussions.
This research was supported by the Armenian HESC through grants
21AG-1C024 (AS), 24RL-1C024 (HT), and 24FP-1F039 (HT, AS).
The authors gratefully acknowledge the computing time made available to them on the high-performance computer Otus at the NHR Center Paderborn (PC²). This center is jointly supported by the Federal Ministry of Research, Technology and Space and the state governments participating in the National High-Performance Computing (NHR) joint funding program (www.nhr-verein.de/en/our-partners).

\bibliography{refs}

\end{document}